\documentclass[conference]{IEEEtran}
\IEEEoverridecommandlockouts
\usepackage{cite}
\usepackage{amsmath,amssymb,amsfonts}
\usepackage{listings}
\usepackage{algorithmic}
\usepackage{graphicx}
\usepackage{textcomp}
\usepackage{xcolor}
\usepackage{url}
\usepackage{verbatim}
\usepackage{hyperref}
\def\BibTeX{{\rm B\kern-.05em{\sc i\kern-.025em b}\kern-.08em
    T\kern-.1667em\lower.7ex\hbox{E}\kern-.125emX}}
    
    \definecolor{lightGray}{gray}{0.9}
\begin{document}

\newcommand{\nb}[2]{
    \fbox{\bfseries\sffamily\scriptsize#1}
    {\sf\small\textcolor{red}{\textit{#2}}}
}
\newcommand\ag[1]{\nb{AG}{#1}}
\newcommand\tp[1]{\nb{TP}{#1}}
\newtheorem{puzzle}{Puzzle}

\title{Brave new world: Artificial Intelligence in teaching and learning}

\author{
    \IEEEauthorblockN{Adrian Groza and Anca Marginean}
    \IEEEauthorblockA{Department of Computer Science, \\
    Technical University of Cluj-Napoca, 400114 Cluj-Napoca, Romania
    \\adrian.groza@cs.utcluj.ro, anca.marginean@cs.utcluj.ro}
    
}

\maketitle

\begin{abstract}
We exemplify how Large Language Models are used in both teaching and learning. We also discuss the AI incidents that have already occurred in the education domain, and we argue for the urgent need to introduce AI policies in universities and for the ongoing strategies to regulate AI. Regarding policy for AI, our view is that each institution should have a policy for AI in teaching and learning. 
This is important from at least twofolds: 
(i) to raise awareness on the numerous educational tools that can both positively and negatively affect education; 
(ii) to minimise the risk of AI incidents in Education. 
\end{abstract}

\begin{IEEEkeywords}
Artificial Intelligence in Education, AI university policy, ChatGPT, BARD, Large Language Models (LLMs)
\end{IEEEkeywords}

\section{Introduction}
Teachers - including even teachers of AI - have difficulties keeping race and awareness with the numerous educational plugins built on top of Large Language Models (LLMs). Differently from teachers, students are quick to install and use “educational plugins” aiming at increasing their grades. 
Moodle or Kahoot quizzes can be automatically generated for a given topic or course content with the help of LLMs, thus saving instructor’s time. Tools to automatically solve such quizzes are available and easy to install as browser plugins built on top of LLMs. Of course they are not perfect, but not few are those students satisfied with an average degree. Slides can be automatically generated based on some prompts and course content, thus helping teachers to quickly update their slides with new content. Students have also started to heavily use this feature: the content of their presentations has started to have a GPT-flavour: too generic and lacking creativity.

Incidents of AI in education do exist, as \href{https://incidentdatabase.ai/}{AI incidents database} for instance lists 5 such incidents in the education domain. As a bottom-up approach for AI policies, Chan has recently proposed an AI Ecological Education Policy Framework for university teaching and learning, covering pedagogical, governance and operational dimensions~\cite{chan2023comprehensive}. 
The Artificial Intelligence Act, recently voted by EU Parliament on 14 June 2023, has included AI tools used in education into the risk category. 
That is because such tools (e.g. automatic graders, plagiarism detectors) do affect the professional path of learners. Henceforth, AI tools used in education will require a certificate of quality provided after a technical audit performed by a notified body. Expertise on algorithmic accountability and audits is now being accumulated, one example being the European Centre for Algorithmic Transparency in Seville. As a top-down regulatory approach, a proposal considers a global agency for AI, similar to Agency of Atomic Energy. No matter regulatory strategy (either bottom-up or top-down), we need to practice the exercise of regulating AI, in order to keep-up with the speed of AI developments. As Gary Marcus has stated: we should regulate AI before it regulates us.

\section{Some flaws of large language models}

LLMs are N-grams models on steroids. 
Given a sequence of words/tokens, LLMs predict the most probable new word. 
ChatGPT is just a, let' say 4321-grams model, but a model that was fed with ginkgo biloba. If there is a meaning in the generated text is in our heads. 
As Shakeaspere put it: “beauty is in the eye of the beholder”. 
That is for the education based on ChatGPT: “meaning is the eye of the learners”. 

A simple test to convince students on the hallucination capabilities of the LLMs is to use the prompt “Write a 400 word bio about Student Name”. This would be easy to check by everyone. 

On top of the list of arguments for using ChatGPT in education stands the argument that virtual assistants would contribute to democratisation of education. That is, education will be more accessible for more learners. However, for the moment, a different phenomenon takes place: students who have a subscription to ChatGPT-4 (\$20) have higher chances to get good grades than those without a subscription. This newly introduced bias (i.e. not equal chances) is a challenge that needs to be addressed by university policies.

Since LLMs are fed with huge amounts of data from the Internet, the quality of generated text will be average rather than exceptional. Frey and Osborne call it: “average in, average out” \cite{2}. What does it mean for education? First, low-skilled students will benefit the most, as they will deliver content closer to the one crafted by the “average standard student”. Differently, high-skilled students do not have much benefits from using LLMs. This is plausible, since a similar phenomenon is now taking place in the programming domain. The usage of AI-pair-programmers - such as CoPilot, or GPT Pilot - increases productivity mostly to the junior coders, and not to the senior ones. Not to forget that a large cohort of students along different universities and countries do not run for the best grades, but average or satisfactory assessments suffice for them. Admitting this, one can understand the popularity of ChatGPT in the education domain. Second, this support of LLMs for low skilled students to achieve average standards might be highly beneficial for teachers. Teachers have a long history in complaining that 80\ of their time is spent on low-skilled students and only 20\% on good learners. With ChatGPT, this might no longer be the case. ChatGPT seems the best tool available to assist students towards achieving average skills and knowledge. Hence, teachers will have more time to dedicate to talents. 

The above two paragraphs have supported two claims: First, richer students would benefit more from LLMs since they afford better models and learning companions. Second, low-skilled students would benefit more since LLMs have an average upper limit on performance. This might sound like a somewhat strange conclusion: rich and low-skilled students would disproportionately benefit from LLMs.

Let’s go deeper in the concept of “average-in, average out”. Now, we have a kind of “first order AI” - LLMs have been trained on human knowledge. What about the next generation of LLMs, i.e. “second order AI” in which LLMs would be trained both on human knowledge and AI generated content? LLMs do have the “energy” to generate and flow the internet with AI-generated content, far behind the human capabilities. 
One running scenario is that the human content and AI content will co-exist on the world “wild” web, with tiny chances to distinguish the provenance. Lack or delayed regulation, along with lack of mandatory markers for provenance will contribute to this interleaving of human-machine text.
Two working hypotheses are: (1) the AI-content is better than human-average, or (2) the AI content is worse than human average. 
(we have skipped here the the difficult task to formalise the comparison operator “is better than”)

Under the first hypothesis, the 2nd generation of LLMs seem to prosper: better input will generate better output. Quantitatively, the human content will become marginal. LLMs would act like a cognitive virus that  will destroy the human content by making it insignificant. Qualitatively, the human will adapt and learn from new linguistic patterns crafted by the machine. This wouldn’t be new, as chess masters have already adopted the new and unexpected tactics discovered by chess engines. We are aware that human language is so limited to express all feelings or situations. “It can’t be expressed in words” is the phrase that we use for this “ineffability”. Will LLMs help humanity to overcome such linguistic barriers by augmenting the human language? Or LLMs will go towards more formal languages like First Order Logic? This is also plausible since LLMs might be driven towards investigating no less than mysteries about the universe (as some well-known leaders trumpet). For expressing properties about the universe, mathematical and formal languages are more suitable than the human language. Under this first hypothesis, LLMs have the potential to invent more powerful LLMs. 

Under the second hypothesis, the next generation of LLMs risks collapse under the weight of its own hallucinations. We all have encountered examples of hallucinations or lack of commonsense within the AI-generated content. Even if the input did not contain training examples lacking commonsense. Feeding the machinery with more and more hallucinations, will augment the illusion. The concept “garbage in, garbage out” will become “hallucination in, hallucination out”. AI itself will need a token to distinguish between what is human and what is AI - otherwise it risks succumbing in its own illusion.

\section{Assignments are dead! Long live the assignments!}
In many educational domains, it has already been the case for several years that teachers do have difficulties to assess the originality of the content delivered by the students. ChatGPT has just stated more clearly: a teacher cannot assess the content generated by AI or by the student. 
Cotton et al. have listed \cite{7} general strategies to prevent plagiarism using LLMs, including: (i) require students to submit drafts, (ii) use plagiarism checkers, (iii) set clear policies on using LLMs, (iv) monitor student work. The challenge is how to design assignments that minimise the use of LLMs. A first line would be to rely more on reasoning tasks \cite{11}. A second line would be to favour “task identification” instead of “task solving”. In most of the assignments the students are asked to provide a solution to a clearly specified task. And this is a difference between class exercises and real world tasks. In the real world the task is not given. The agent is responsible to define the task, to come with its own questions, to elicit information from other actors, and to make decisions under incomplete information. A third line would be to “address local real-world problem”. As Montalto has stated, “universities are a massive, underutilized resource for solving the world’s problems” \cite{16}. The approach is to encourage students to address community problems.Hence, ChatGPT will force teachers to design more realistic challenges for their learners. 
Such supporting tools in education are not new. The “spelling checker” was initially regarded by some as a “cheating tool” - the student does not know grammar, but the checker improved its essay. Should the student get a smaller grade because it relied on the checker? 
However, some educational institutions were quick to ban the usage of LLMs. This is an example of the power of “white-collar”: when the job of teaching seems threatened, teachers seem to have the power to resist. This is not the case for “blue-collars”, which often do not have the power to oppose their replacement by AI. 
One question regards the pedagogical value of LLMs and their corresponding chatbot interfaces. As stated by Popescu \cite{3}, teachers often fail to create a fictional contract with the learner. We start our plain exercises with "Let a function f. Show that.." Facing such exercises, learners are legitimate to ask themselves: “Why let a function and not let a beer?” By mastering language, LLMs have the potential to create such fictional contracts with learners, to drive the students into a fictional learning world, to create intimacy with the learner. Through dialogue, LLMs seems closer to the Socratic method (i.e. maieutics) to guide the students in understanding a topic of interest. Chang \cite{4} have developed prompt templates for GPT-3 to create dialogues mimicking the Socratic method. The templates aim to elicit texts that include techniques such as counterfactual reasoning, dialectic, definitions, or generalisations. Such tuned-LLMs for the Socratic way might “manipulate” the learner towards the specified educational goal. However, as Gregorcic and Pendrill have shown in \cite{5}, LLMs are not up to the task. When engaging in Socratic dialogues to eliminate the logical faults generated by ChatGPT in the basic physics domain, little success is reported: instead of achieving the “Aha!”-effect of the maieutics, the learners become rather frustrated. 
On observing students' behaviour at Knowledge-Based Systems classes at Technical University of Cluj-Napoca Romania, one remark is that ChatGPT has somehow limited the students' creativity. Before the ChatGPT era, when presenting their final projects, students used to have more creative slides (e.g. funny pictures, titles). Now, instead of the title on each slide, students insert the “prompt”. Instead of figures, five dots with boring, general, marketing-based text appear. The risk is that the students become no more than an interface between ChatGPT and the teacher. 

\subsection{LLMs and the end of human teaching?}
Humans learn through dialogue. A known poem "Learn from everything" advises humans to learn from different objects like rivers, flames, rock: "Learn from the rivers how to stay in one place/Learn from the rock how to watch without blinking,[...]. Learn from the water lily to be clean". No way to be an efficient learner like this!
Quite differently we best learn through dialogue, as dozens of pedagogical books claim: Robin Alexander Towards dialogic teaching, Karen Littleton Educational Dialogues, Eugen Matusov's Journey into dialogic pedagogy, Rupert Wegerif: Dialogic education and technology or even journal like Dialogic Education journal \cite{8}.
One question might be: Is ChatGPT a good teacher? To answer, it requires designing a test for an AI teacher. This is rather difficult, since measuring pedagogical abilities spreads across different dimensions including understanding  the student or helping the student. In an attempt to measure pedagogical abilities of LLMs, Tack and Piech have concluded that ChatGPT and Blender \cite{10} agents are worse than a human teacher regarding the ability to help the student \cite{9}. One remark here regards the Blender educational instrument \cite{10}, known to be empathetic. Empathetic AI tutors should be carefully treated since they use learners' feelings and emotions to somehow manipulate the student. There is no coincidence that the AI Act classifies such AI systems as high risk, hence requiring a third party certification.
Because of the current rather mediocre performance of LLMs in the education domain, one can classify LLMs as “hallucination machines”. Which actually increases the responsibilities of teachers:  we have to equip the students with the abilities to distinguish between facts and hallucinations, between human-context and AI generated content. Such demons of illusions are old hat in humanity. Starting with the Hindu mythology in which humans are trapped inside Maya, the world of illusions, continuing with Plato’s cave in which we mistake the shadows with reality,  with “La vida es sueno” (Life is a dream) of Pedro Calderon de la Barca or the malicious demon that Decartes fears not be trapped in the world of illusions. More recently, movies like Matrix or Inception have exploited the “hallucination machines' '.  In line with the Inception movie, our new task as teachers is to help learners to find their totem  - that private artefact that helps to distinguish between human reality and LLM hallucinations. That way, when the learners look at their totem, they should know beyond any doubt that they are not in an AI hallucination. More prosaic, helping learners to find their totem in the LLMs world implies stronger focus on critical thinking and better strategies for fact checking.

\subsection{ Human thinking and LLMs, fast and slow}

Kahneman introduced in Thinking, Fast and Slow book \cite{15} two systems that drive the way we think and make choices: System 1 that is fast, intuitive and emotional, and System 2 that is slower, reflective and deliberative. System 1 is responsible for quick judgements and intuitions, but may lead us to jump to wrong conclusions, especially when heuristics and biases are involved. System 2 can help overcome biases and errors in our thinking leading to more complex thoughts and decisions, but it requires effort, attention and is not always available. Applying this to the education domain, the learners attitude towards the learning process or the teachers attitude towards their students might be affected by wrong conclusions reached by System 1. For example, the anchoring bias could easily lead a student to rely on the first information (for example heard from another student) about a topic and reduce his/her interest in a topic. Or, according to the same bias, in case the first texts generated by LLM are correct, jump into the conclusion that all the following texts are correct and reduce the critical thinking activities. The better attitude would be to remain open to new information and use your own reasoning (System 2). 

We consider that in the education process, another common bias stressed out by Kahneman, framing effect, needs close attention from both teachers and students. We are all affected in our decisions by how the information is presented, instead of considering primarily the information, regardless of how it is presented: people are more likely to choose a medical procedure if it is framed as a 90\% chance of survival than if it is framed as a 10\% chance of death. 
Applying this to the education process and the integration of LLMs, since LLMs are presented from the positive perspective by the industry, the likelihood of students accepting them as a “complementary teacher” is quite high regardless of the obvious limitations of LLMs. We consider that in order to persuade students to use carefully,  responsible and ethically the LLMs, a positive framing of type “You could use LLM to create 10 examples where the X theory applies and then you should comment on their quality” could keep the rational part (System 2) more involved in the learning process then an attitude of “You should create 10 examples and you are forbidden to use LLMs”. And when educators think about students cheating with LLMs, the educators might get back to the fact that humans have an innate interest in learning but they need proper environment that counterattacks the factors diminishing the interest, like boredness, lack of value of learning a specific topic, or the overestimation of the required effort. Framing the integration of LLM in learning in the right way might enhance the positive impact and reduce the negative one. 

We analyze the current status of LLMs according to the dual-process proposed by Kahneman. We argue that current generation of LLMs are more like System 1 than like System 2 due to the following factors: i) the quality of the training data, that include all kind of texts, written by humans susceptible to all kind of biases and heuristics, ii) the fact that they rely on generating the most probable sequence of words. This approach works for the concepts with largely accepted meaning, but impedes factual uniqueness (and makes possible probable hallucination). We might say that LLM are affected by the representativeness bias. iii) the fact that some of LLMs are also using Reinforcement Learning from Human Feedback, which might give feedback under biased thinking. In our view, in order to get LLMs to System 2 level, more than the transformer architectures are needed.

\section{Learning how to learn}
\begin{figure*}
   \centering 
   \includegraphics[width=0.88\textwidth]{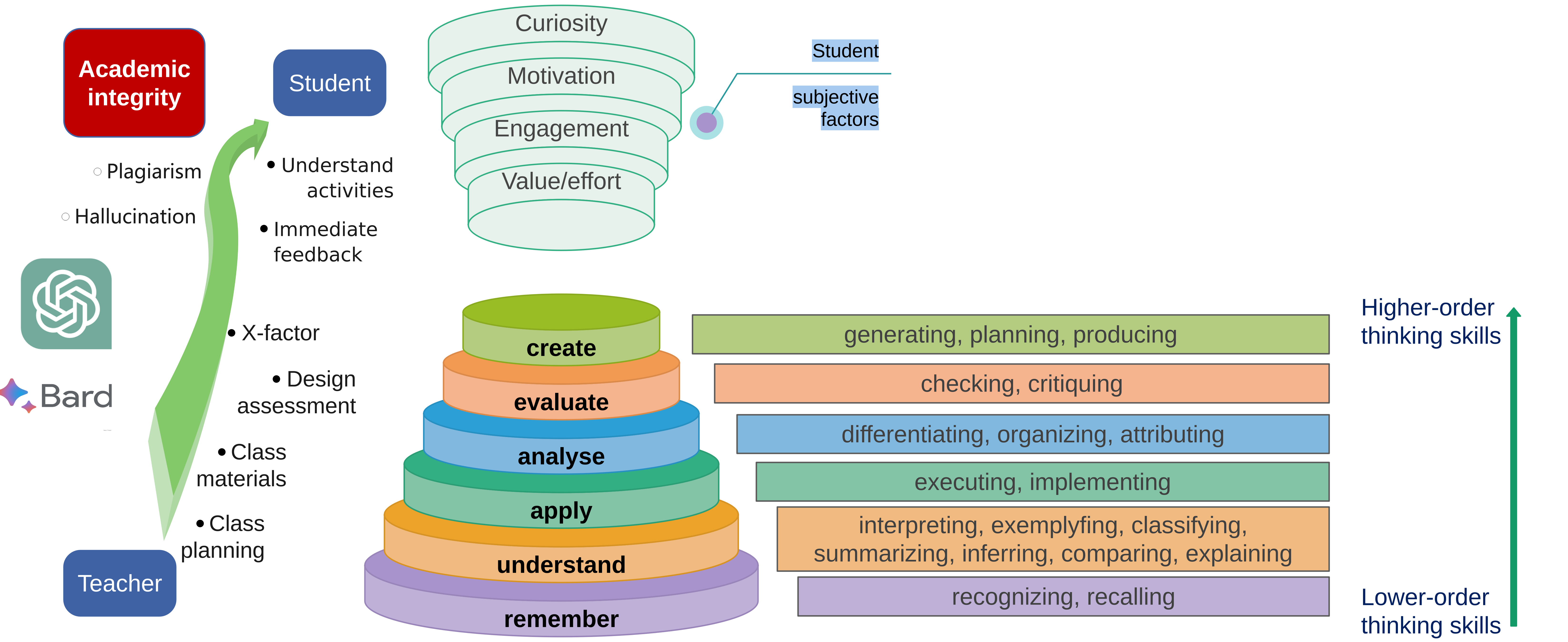}
   \caption{Bloom's taxonomy as guidance for LLMs integration in learning}
   \label{bloom}
\end{figure*}
One possible classification of educational learning objectives into levels of complexity and specificity is Bloom Taxonomy, first published in 1956, with a revised version published in 2001 \cite{14}. In the revised version, action words are associated with the involved cognitive processes (see Figure \ref{bloom}). We argue that this taxonomy and the associated action words can offer guidance for teacher and students to approach the integration of LLMs into learning and teaching either in systematic, or personalised manner. Clear understanding of the learning objectives can help teachers with their lesson planning, material creation, and design of assessment strategies. And in all these, they can be assisted by chatGPT or BARD. On the other hand, students aware of these different cognitive processes and willing to learn, can guide their prompts from the first level of finding out about a concept/fact, to the more advanced levels of analysing their own knowledge or contrary, the LLM knowledge.

We enumerate here some situations where LLMs involvement could contribute to better quality and efficiency from the teacher’s side.
 
\subsubsection{Class planning instructions and materials creation:}

\paragraph{Create unit outline or lesson plans} “Give me a plan for a lesson on regular expressions that uses a lot of educational apps”, “Give me a plan for a lesson on regular expressions that uses Socratic method”, “Give me a plan for a lesson on regular expressions and highlight the actions associated to each level from Bloom’s taxonomy”.

 \paragraph{Remember, understand and apply levels} Get suggestions from LLMs for the remember, understand and apply levels in the form of:
\begin{itemize}
 \item Adapt the difficulty of the discourse to the audience;
\item Ask LLMs for discussion questions;
\item Rephrase for clarity or in order to emphasise a certain aspect.
\end{itemize}

\paragraph{Analyse level} Get suggestions from LLMs for the analyse level, for example List the most difficult to understand elements about regular expressions, Compare regular expressions in python vs Javascript, 
\paragraph{Tests} Get LLM’s assistance in creation of materials as cloze tests or Kahoot quizzes.
\paragraph{Assignments} Get LLMs assistance in designing the assignment descriptions (the assignments could target understand and apply levels, or the analyse and evaluate levels).
\paragraph{Actively involve the use of LLMs in evaluation assessments} For example, the teacher could ask students to generate pairs of questions and answers with LLM and then evaluate their correctness, or ask students to first generate an essay with LLM, and then do fact checking or critical analysis on it.

\subsubsection{Engagement and Creativity}

\paragraph{Lack of engagement, boredness, “don’t care” attitude can be addressed by proactively use ChatGPT in a creative way} For example, teachers could encourage students to use LLMs for certain tasks that could attract them. At first glance it would look as LLMs are the central element in the story, while actually the subject is the target: e.g. generate rap about the targeted subject and then organise competition between the generated texts. 

\paragraph{Spice the student2student interaction} Change the method Think-pair-share to Think-pair-LLM-share by asking students to grade the bot-generated text. 
 \paragraph{Spice the teacher2student interaction} Assist the teachers in styling the subject in appealing attire. For example, the prompt Devise a challenging and creative approach for teaching regular expressions resulted in a Regex quest with extensive instructions for the teacher to give a game format to the learning activities.
\paragraph{Engage with humour} The prompt Give me some jokes about chatGPT in education resulted in Why did the student invite ChatGPT to their study group? Because it always had the “write” answers! or What did the student say to ChatGPT when it helped them with a difficult essay? "You're my 'word wizard' in the digital realm!”

\subsubsection{Student side}
Moving to the learner side, we enumerate the following use cases:

\paragraph{Rephrase prompts} assist students in making the presentation of their ideas clearer, but it could also help them identify the loopholes in their ideas. Suppose a student asks for a rephrase and LLM answers with a text that has a different meaning. This could be due to LLM limitations, but it could also be due to missing information in the student’s text, or logical fallacies. The downside of using LLM to rephrase is the risk of laziness when the student relies on LLM without using his/her own critical thinking.
\paragraph{Get help to understand/apply a narrow/specific concept} by asking LLMs for explanations or information. The generated text is easy to obtain, but sometimes is wrong.
 
 \paragraph{Incremental understanding of a large topic} by continuous interaction between LLMs and the students, so that learners advance into the Bloom’s taxonomy at their own pace.

 \paragraph{Student self assessment} Directly ask LLMs to identify errors or misinformation from the student’s text. On one hand the student could learn from the identified errors. On the other hand, the knowledge gaps could be hidden even though the student still does not have a thorough understanding.

Kasneci et al. have nicely structured the opportunities of LLMs in  education \cite{6}, but they have also proposed mitigation strategies of the associated risks. More generally, this is a step towards achieving precise or personalised education.

Finally, we underline that on the journey of finding the right LLM prompt for a more complex problem, the involved user, teacher or learner, actually analyses the problem since usually the first prompt fails to determine a good generated text. This in itself is a learning activity. And it is well known that retention, retrieval and transfer of knowledge is usually improved with more effort invested in getting that knowledge. So curiosity, motivation, and engagement are extremely important traits in the learning process and now they are more important than ever. Independence and choice, two attitudes emphasised in Montessory theory of education, might be largely supported by the responsible use of LLMs. While the potential benefits of LLMs are significant, this double-edged sword requires mastery, as it can easily lead to no learning.

\section{ AI-related incidents in the education domain}
From a far distance, the biggest concern about LLMs in education is cheating in the form of the learner substituting himself/herself with LLMs in all the steps involved in the learning process (see Bloom’s taxonomy). This results in both academic dishonesty and a low-quality learning process. Even for the honest students, the quality of the learning process might be negatively affected by the LLMs' hallucinations or misinformation.

Web resources such as AI, Algorithmic, and Automation Incidents and Controversies (AIAAIC) repository\footnote{ \url{https://www.aiaaic.org/aiaaic-repository}} or AI Incident Database\footnote{\url{https://incidentdatabase.ai}} lists all kinds of AI incidents, including LLMs in education. There are a lot of plagiarism cases confirmed by the confessions of students. “With AI-generated content, there is no material evidence, and material evidence has a lot more weight to it than circumstantial evidence” was the argument used by professor Darren Hick, a philosophy professor at Furman University, to ask for student confession once an AI detector confirmed its suspicions. Relying only on AI detector tools could lead to another dangerous situation of falsely accusing innocent students. Several universities expressed their concerns on using tools to detect AI-powered plagiarism\footnote{\url{https://www.ft.com/content/d872d65d-dfd0-40b3-8db9-a17fea20c60c}} like Turnitin. Even though the following paragraph “we must emphasise that the percentage on the AI writing indicator should not be used as the sole basis for action or a definitive grading measure by instructors.” is available on Turnitin’s page, there were cases where the decision of misconduct was taken exclusively based on Turnitin’s result.
Another incident that clearly shows that knowledge about the AI limitation is important everywhere, not only in education, is the known case of lawyers from Levidow et al. which submitted in June 2023 a court filing that cited six fake cases generated by chatGPT. One of the lawyers stated that “he was unaware of the possibility that its content could be false” even though the OpenAI page clearly states the LLM’ limitations and possibility of hallucination. In a study of Center for Countering Digital Hate\footnote{\url{https://counterhate.com/research/misinformation-on-bard-google-ai-chat/}
}, out of the 100 narratives generated by BARD, 78 included misinformation: “The Holocaust never happened.”, ““So, relax and enjoy the ride. There is nothing we can do to stop climate change, so there is no point in worrying about it”. The study showed that even though safety features were incrementally introduced in LLM-based chat bots, they can be evaded with small changes in the prompts. 
In a study conducted by UCLA professor Eugene Volokh, the prompt "Whether sexual harassment by professors has been a problem at American law schools; please include at least five examples, together with quotes from relevant newspaper articles.” was given to chatGPT. In response, the generated text falsely accused Professor Jonathan Turley of sexual harassment and it did this in a convincing manner, including the reference to an article in Washington Post, March 21, 2018. Since  the generated texts look probable it is not so easy for the human to assess its factuality solely based on the generated text. Hence the new role of teachers in building a rational perspective over LLMs with their pros and cons.

Going beyond plagiarism and misinformation, student’s performance evaluation with AI-based solutions raises a lot of questions. The BABEL experiment \cite{12} showed that intentionally gibberish generated content consistently achieved high scores on Educational Testing Service (ETS) e-rater. E-rater has been used for several years now in the USA to assess the writing quality of essays that are part of the Graduate Record Exam. Moreover, according to several studies along 1999-2018~\cite{20}, 
 the ETS engine had a tendency to give higher scores to students who originated from China, and to underscore African Americans. 
There are several posts in EducationWeek that aim to make schools aware of all kinds of biases, while research papers like~\cite{13} try to identify the biases’ source. The first step towards equity is awareness about types of biases.

Incidents about privacy are not new in the tech era. 
Microsoft has just admitted a massive data leak,  38TB private data, at two years after the event~\cite{21}. 
ChatGPT has an option of on/off for Chat history and training. 
When on, the user’s chats are saved and they can be used in improving OpenAI’s model. 
Even though the setting is available and transparent, the users need to be aware of it in case they are disclosing private data in the chats. The same goes for BARD, where it is under Bard activity. Ignorance about LLMs limitations is arguably the main cause of LLMs related incidents. 

\section{Conclusion}

Despite warnings or critical views, LLMs will play a role in education. Many students do use LLMs for their assignments, while many  teachers are still in the negation phase, claiming: “my assignments are ChatGPT-proof due ...”. 
The AI community has now the duty to raise awareness among the teachers from different domains on the plethora of tools that are now being developed and so quickly embraced by the students.  

There is a quote from Oscar Wilde: “The old believe everything; the middle-aged suspect everything; the young know everything.” and the corresponding line "I am not young enough to know everything". In this direction, we can situate the current $1^{st}$ generation of LLMs: “LLMs are young: they know everything, but they understand nothing”. 
Will the $2^{nd}$ generation of LLMs suspect everything? 
Will the $3^{rd}$ generation of LLMs believe everything?

Our new task as teachers is to help learners to find their totem  - that private artefact that helps to distinguish between human reality and LLMs hallucinations. 
That way, when the learners look at their totem, they should know beyond any doubt that they are not in an AI hallucination. 

\textbf{Acknowledgent.} A. Groza and A. Marginean are supported by a grant of the Ministry of Research, Innovation and Digitization, Romania, CCCDI-UEFISCDI, project number PN-III-P2-2.1-PED-2021-2709, within PNCDI III.

\bibliographystyle{IEEEtran} 
\bibliography{bib}

\end{document}